\documentclass[%
reprint,
 amsmath,amssymb,
 aps,
pra,
]{revtex4-1}

\usepackage{graphicx}
\usepackage{dcolumn}
\usepackage{bm}

\begin{document}


\title{On random-matrix bases, ghost imaging and x-ray phase contrast computational ghost imaging}
\author{David Ceddia}
\email{David.Ceddia@gmail.com}
\author{David M. Paganin}
\affiliation{School of Physics and Astronomy, Monash University, Victoria 3800, Australia}
\date{\today}

\begin{abstract}
A theory of random-matrix bases is presented, including expressions for orthogonality, completeness and the random-matrix synthesis of arbitrary matrices. This is applied to ghost imaging as the realization of a random-basis reconstruction, including an expression for the resulting signal-to-noise ratio.  Analysis of conventional direct imaging and ghost imaging leads to a criterion which, when satisfied, implies reduced dose for computational ghost imaging. We also propose an experiment for x-ray phase contrast computational ghost imaging, which enables differential phase contrast to be achieved in an x-ray ghost imaging context. We give a numerically robust solution to the associated inverse problem of decoding differential phase contrast x-ray ghost images, to yield a quantitative map of the projected thickness of the sample.  
\end{abstract}


\maketitle

\section{Introduction}

Conventional direct imaging is typically realized using position-sensitive detectors such as those that are currently used in digital photography, x-ray radiography, neutron imaging, electron microscopy {\em etc}. Ghost imaging \cite{erkmen2010ghost, padgett2017introduction}  is an emerging alternative in which the conventional direct-imaging methodology is replaced with an indirect-imaging process which synthesizes an image given a series of known illuminations and associated position-insensitive ``bucket'' detector measurements. 

In its simplest form, ghost imaging synthesizes an image from a set of temporally acquired single-pixel detector (bucket reading) outputs which are each correlated with a random intensity pattern.  Explicitly:  
\begin{align} \label{eq: Ghost Image Formula}
\text{Image} = \frac{1}{N} \sum_{k=1}^{N} \left( B_k - \text{E} \left[B \right]  \right) I_k(i,j),
\end{align}
where $N$ is the number of spatially-random intensity patterns $I_k(i,j)$ which is indexed by $k$ and $\text{E} \left[B \right] $ is the expectation value of the bucket reading:  
\begin{align}
B_k = \sum_{i=1}^{n} \sum_{j=1}^{m} f(i,j) I_k(i,j) =\left< f,I_k \right>.
\end{align}
Here, $\left< \cdot \right> = \sum_{i=1}^{n} \sum_{j=1}^{m} $, $f(i,j)$ is the sample intensity pattern and $(i,j)$ denote the pixel coordinates \cite{bromberg2009ghost}. 

Ghost imaging originated in visible-light studies \cite{Klyshko1988,Belinsk1994,strekalov1995observation,pittman1995optical,bennink2002two,scarcelli2006can,kolobov2007quantum,shih2012physics}.  The field has been advanced by incorporating compressive sensing ideas \cite{katz2009compressive, zerom2011entangled, katkovnik2012compressive, yu2014adaptive} and studies of turbulent environmental robustness such as lidar systems \cite{zhao2012ghost}, atmospheric turbulence in thermal imaging \cite{shi2012adaptive} and underwater fluctuations in optical imaging \cite{le2017underwater}. Visible-light ghost imaging has been augmented to achieve absorption-contrast ghost imaging using x-rays \cite{Yu2016,pelliccia2016experimental, schori2017,zhang2018tabletop, pelliccia2017practical} and atoms \cite{Khakimov2016}.  The potential for reduced dose \cite{zhang2018tabletop}, increased resolution, compressive sensing capabilities \cite{katz2009compressive} and turbulence robustness \cite{turbulence1} are all under ongoing investigation. 

The underlying concept of ghost imaging has been pointedly distilled to ``a vector projection of the [sample] transmission function over [N] different random vectors'' \cite{katz2009compressive}, or ``a form of orthogonal [random] function expansion'' \cite{pelliccia2017practical}. We build on these ideas to develop a mathematical theory of random-matrix bases, with particular focus on the inherent noise associated with such a reconstruction.  We develop an inequality which, if satisfied, implies that computational ghost imaging exhibits reduced dose when compared to its direct imaging counterpart.  If this inequality is not satisfied, this implies that alternative reconstruction strategies---including but not limited to those based on the concept of compressive sensing---are necessary for ghost imaging to reduce dose in comparison to direct imaging.   

Owing to the potential for reduced dose \cite{zhang2018tabletop}, a promising application of ghost imaging is in x-ray medical imaging. As previously mentioned, already realized is experimental set-ups for x-ray absorption contrast ghost imaging. Outstanding, however, is a means to realize x-ray phase contrast ghost imaging. In seeking a means to achieve this, we are motivated by the significant advances in x-ray direct imaging that have been enabled through the use of phase contrast \cite{Wilkins2014}.  There exist means for phase contrast ghost imaging in the optical regime \cite{shirai2011ghost}, although the experimental set-ups are not readily transferable to the x-ray regime. We propose an experimental set-up to achieve x-ray differential phase contrast computational ghost imaging (PCCGI).
We also consider the associated inverse problem, of inverting the differential phase contrast x-ray ghost image to obtain a quantitative map of the projected thickness of the sample which led to such an image.    

We close this introduction with an outline of the remainder of the paper.  Section II develops the formalism of using a set of real, random $n \times m$ matrices as a basis over which arbitrary $n \times m$ matrices may be decomposed. While we view such decompositions in the context of imaging, where the matrix to be decomposed is a pixelated image, the formalism may be applied to a broader range of discretized fields. Loosely, the procedure may be viewed as synthesizing signals by superposing noise. The statistical orthogonality of the random matrices is treated, together with completeness and an error metric for the statistical errors in the random-matrix decomposition. A comparison is also drawn between directly decomposing a given pixelated image using a random-matrix basis, and decomposition using a basis that has been orthonormalized using the Gram--Schmidt process.  Section III gives a comparison between direct and computational ghost imaging with respect to fluence/dose. An inequality is developed which, if satisfied, implies a particular ghost imaging scenario to have reduced dose relative to its direct imaging counterpart.  Section IV proposes a means for x-ray phase contrast computational ghost imaging, namely a means for x-ray computational ghost imaging that is sensitive to the phase of the coherent x-ray field that impinges on the specified x-ray ghost imaging system.  This gives a technique for x-ray ghost imaging that is sensitive to the refractive properties of a sample.  A solution is developed to the associated inverse problem of quantitative phase--amplitude reconstruction of the projected thickness of a single-material sample, using x-ray phase contrast ghost imaging.  Section V discusses the broader significance of some of our results, together with some future avenues for research.  We conclude with Section VI.

\section{Random-Matrix Bases}

The ``Infinite Monkey Theorem'' \cite{Eddington1928} states that an infinite random sequence almost surely---{\em i.e.} with a probability that is infinitesimally close to unity---contains every possible finite subsequence. In principle, this permits synthesis of an arbitrary sequence via one realization of a random sequence.  To say that such a process is extremely inefficient would be an understatement, owing to the impracticably large mean times one needs to wait for a given desired sequence to occur. In an alternative, practical approach, a given discrete random output can be cut into pieces of the size of the desired sequence to form a random basis. Each random basis member can then be superposed in a weighted sum to synthesize a desired signal (or some approximation of it). Given that the present paper works in an imaging context, with particular emphasis on computational imaging and ghost imaging, we will discuss the topic of random bases in the context of a matrix basis. Here, each element of the matrix may be associated with a pixel in a Cartesian lattice.

Consider the standard matrix basis set $\{e_{ij}\}$ that spans $M_{nm}$ ({\em i.e.} $nm$ distinct matrices of size $n \times m$, each of which has a single element equaling unity, with all other elements being zero). We will define a random basis member as corresponding to the randomly weighted sum of the standard set:
\begin{align}
R_{k} = R(1,1) e_{11} + ... +  R(n,m) e_{nm}.
\end{align}
Here, $k \in [1,N]$, $i \in [1,n] $, $j \in [1,m]$ and $R(i,j)$ is the realization of a zero centered random variable $R$ drawn from the probability density function Pr$(R)$. We reserve defining an explicit probability density function for $R$ ({\em e.g.} uniform, Gaussian, Poissonian, {\em etc}.) and instead leave our results in terms of parameters of that distribution ({\em i.e.} expectation value E$[R]$ and variance Var$[R]$). 

Alternatively, we can work in reverse from the random bases via the definitions:
\begin{align}
\begin{split}
\text{E} [R] &= \lim_{n,m \to \infty} \dfrac{1}{nm} \sum_{i=1}^{n} \sum_{j=1}^{m} R_k (i,j) \\
&= \lim_{N \to \infty} \dfrac{1}{N} \sum_{k=1}^{N} R_k(i,j),
\end{split}
\end{align}
and
\begin{align}
\begin{split}
\text{Var}[R] &= \lim_{n,m \to \infty} \dfrac{1}{nm} \sum_{i=1}^{n} \sum_{j=1}^{m} (R_k (i,j) - \text{E}[R])^2 \\
&= \lim_{N \to \infty} \dfrac{1}{N} \sum_{k=1}^{N} (R_k (i,j) - \text{E}[R])^2
\end{split}
\end{align}
where these equalities do not necessarily hold for any finite choice of $n,m,N$ but are increasingly well approximated with increasing $n,m,N$. Random bases defined in this way will display ergodic behavior in expectation.  Ergodicity is usually defined as the equality of ensemble and time averages \cite{mandel1995optical}.  The related but different concept of ``ergodicity in expectation'' refers to the expectation value over a matrix set being equal to the expectation value over the spatial distribution, where the statistics of each will be governed by Pr$(R)$.   
That does not mean, however, that the spatial average of each realized basis member will be the same, nor will each set average for a given coordinate $(i,j)$ be the same. In each case these are finite sums that are ultimately subject to variance.

We consider two related cases: (i) synthesis of a desired matrix $f(i,j)$ using a non-orthogonal discrete random-matrix basis and (ii) synthesis using an orthonormal discrete random basis. The first case is relevant {\em e.g.} to ghost imaging using spatially-random shot-noise fields, with the latter case being relevant {\em e.g.} for computational imaging using specifically designed pseudo-random masks. Owing to the generation of a particular $N$-member random-matrix basis being a stochastic process, a probabilistic approach is employed. 
For the non-orthogonal discrete random-basis case, the exact function is retrieved in expectation and noise is quantified in the variance. The orthonormal random basis is achieved via applying the Gram--Schmidt process \cite{AntonRorres, Diaconis} to a given non-orthogonal set of random matrices and forms a more efficient reconstruction method when compared to the first case listed above. It is important to note regarding a particular realized set of $N$ random matrices, each real element of which is by assumption generated by the same specified probability distribution, that subsequent Gram--Schmidt orthogonalization yields a sequence of random matrices whose statistics in general change with position in the matrix sequence. 
This implies a slightly different approach for the two cases.

\subsection{Orthogonality} \label{sec: expected orthog}

We assess orthogonality of two different random matrices by considering their Frobenius inner product \cite{Horn}, namely the discrete version of the inner product between two real functions of two real variables:
\begin{align} \label{eq: inner product orthogonality}
\left< R_k, R_{k'} \right> = \sum_{i=1}^{n} \sum_{j=1}^{m}  R_k(i,j)  R_{k'}(i,j).
\end{align}
We restrict ourselves to the case E$[R]=0$ ({\em i.e.} the distribution has been zero centered, which can be obtained from an arbitrary random variable via $R \equiv X - \text{E}[X]$). Being a random process, the probability of these basis members being strictly orthogonal is typically of zero measure. 
Despite not being able to make deterministic statements regarding orthogonality, we can make probabilistic statements regarding expectation value and variance of orthogonality. To calculate the expected value, consider:
\begin{align}
\text{E} \left[ \left< R_k, R_{k'} \right>  \right] 
&= nm \ \text{E} \left[   R_k(i,j)  R_{k'}(i,j) \right] \nonumber \\
&= nm \ \text{Var}[R] \delta_{k k'},
\end{align}
where $\delta_{k k'}$ is an ensemble Kronecker delta (1 for $k = k'$, and 0 for $k \neq k'$). From this result, we can see that regardless of the particular distribution of $R$, so long as $R$ is zero centered, two random basis members will be orthogonal in expectation. 
We emphasize the difference between ``orthogonal'' and ``orthogonal in expectation'': the former label refers to two random matrices having zero Frobenius product, with the latter stating that the Frobenius product is a random variable with zero expectation value. 

We turn to the deviations from orthogonality that we can expect as quantified by the variance. The variance of the Frobenius product expressed in Eq.~(\ref{eq: inner product orthogonality}) is:
\begin{align}
\text{Var} \left[ \left< R_k, R_{k'} \right> \right] &= nm \ \text{Var} \bigg[   R_k R_{k'}  \bigg]  \nonumber \\
&= \begin{cases}
nm \ \text{Var} \left[   R  \right]^2 & \text{for } k \neq k' \\
nm \ \text{Var} \left[   R^2  \right] & \text{for } k = k',
\end{cases} \label{eq: var orth}
\end{align}
where we have used the independence of each random number realization to set the covariance to zero. 
Note also, that although $R_k$ and $R_{k'}$ are different random numbers, their expectation value and variance will be the same since they are drawn from the same distribution. The larger the matrix size $n \times m$, the larger the variance from orthogonality. Although, if we were to normalize the inner product with $1/(nm$Var$[R])$, the right hand side of Eq.~(\ref{eq: var orth}) would decrease with increasing matrix size.  We note, in this context, that our normalization differs from that which is typically used in the theory of random matrices \cite{RandomMatrixBook}. 

\subsection{Completeness relation} \label{sec: completeness}

We turn attention to the completeness relation:
\begin{align} \label{eq: Completeness Relation}
\begin{split}
\lim_{N\to \infty} \frac{1}{N \ \text{Var}\left[ R \right] } \sum_{k=1}^{N} R_k(i,j)  R_k(i',j') \\
=  \delta (i-i',j-j'),
\end{split}
\end{align}
where $\delta (i-{i'},j-{j'})$ is a spatial Kronecker delta and all other symbols are as previously defined. We are particularly interested in the finite-$N$ version, namely:
\begin{align} 
\begin{split}
\frac{1}{N \ \text{Var}[R]} \sum_{k=1}^{N} R_k(i,j) R_k(i',j') 
\approx \delta (i-i',j-j'),
\end{split}
\end{align}
where of interest is the rate at which the approximation will converge to a Kronecker delta, as quantified by the variance of the above sum. Confirming we indeed obtain the Kronecker delta as claimed, consider the expectation:
\begin{align}
\begin{split}
\text{E} \left[ \frac{1}{N \ \text{Var}[R]} \sum_{k=1}^{N}R_k(i,j) R_k(i',j') \right] \quad\quad\quad\quad\quad\quad \\
= \frac{1}{\text{Var}[R]}  \text{E} \left[  R(i,j) R(i',j') \right] =  \delta (i-i',j-j'),
\end{split}
 \end{align}
where we have used the fact $R$ is a zero centered random variable, which implies that $\text{E}[R^2] = \text{Var}[R]$. So, we indeed obtain a spatial Kronecker delta as an expectation. 

Moving onto the variance, this can be calculated via:
\begin{align}
\begin{split}
\text{Var} \left[ \frac{1}{N \ \text{Var}[R]} \sum_{k=1}^{N}  R_k(i,j) R_k(i',j') \right] \\
= \frac{1}{N^2 \ \text{Var}[R]^2} \sum_{k=1}^{N} \text{Var} \left[  R(i,j) R(i',j') \right]
\end{split}  \nonumber \\
&= \begin{cases}
\dfrac{1}{N} & \text{for } (i,j) \neq (i',j') \\
\vspace{-0.4cm} \\
\dfrac{1}{N} \dfrac{ \text{Var} \left[ R^2 \right]}{\text{Var} \left[ R \right]^2}  & \text{for } (i,j) = (i',j'),
\end{cases} \label{eq: var comp}
\end{align}
where we have again used the fact that each of the covariances is zero since each $R_k(i,j)$ is independent. As the number $N$ of basis members is increased, the closer we converge to the expectation value of the spatial Kronecker delta.  As $N \rightarrow \infty $, we achieve a complete (indeed, over-complete) set of non-orthogonal basis members.

\subsection{Synthesis of arbitrary matrices using a random-matrix basis} \label{sec:Derivation of Fn Rec w/ RB}
Suppose one wishes to express an $n \times m$ arbitrary discretized function or matrix $f(i,j)$ as a linear combination of realizations of random $n \times m$ matrices, each element of each random matrix being an independent deviate drawn from the same arbitrary real probability distribution. To synthesize $f(i,j)$, take the completeness relation expressed in Eq.~(\ref{eq: Completeness Relation}), multiply both sides by $f(i,j)$ and then sum over all spatial points to obtain:
\begin{align}
\begin{split}
f(i,j) =  \sum_{{i'}=1}^{n} \sum_{{j'}=1}^{m} f(i',j')~ \delta (i-i',j-j') \quad\quad\quad\quad\quad\quad\quad \\
= \lim_{N\to \infty} \frac{1}{N \ \text{Var} \left[ R \right]} \sum_{{i'}=1}^{n} \sum_{{j'}=1}^{m} f(i',j')  \sum_{k=1}^{N} R_k(i,j)  R_k(i',j').
\end{split}
\end{align}
Interchanging the order of the sums and approximating the infinite sum with a finite sum, we see that $f(i,j)$ is approximated with finitely many terms:
\begin{align} \label{eq: function reconstruction non-orthog}
f(i,j) 
& \approx \frac{1}{N \ \text{Var} \left[ R \right]} \sum_{k=1}^{N}  w_k R_k(i,j) \equiv f_{(N)}(i,j),
\end{align}
where $w_k \equiv \left< f, R_k \right>$ is the weighting coefficient of the random matrix $R_k(i,j)$  and $f_{(N)}(i,j)$ is the $N^{\text{th}}$ order approximation to the exact expression $f(i,j)$. This is a key expression, for it demonstrates that a signal $f(i,j)$ can indeed be approximated by a linear combination of noise maps.  A decomposition of this nature is commonly employed in computational imaging and ghost imaging \cite{bromberg2009ghost,shapiro2008computational}, although the preceding analysis is not tied to either particular context.   

To quantify the noise in a reconstruction made with a finite number of basis elements, we calculate the variance:
\begin{align}
\begin{split}
\text{Var}  \left[  f_{(N)}(i,j) \right] 
& =  \frac{1}{N}\sum_{{i'}(\neq i)=1 }^{n} \sum_{{j'}(\neq j)=1}^{m} f^2(i',j')  \\ & + \frac{f^2(i,j)}{N} \left( \frac{\text{Var}\left[R^2 \right]}{\text{Var} \left[ R \right]^2}\right),
\end{split}
\end{align}
where we have twice used that each $R_k(i,j)$ is independent and the covariance of each term with respect to the others is zero. For $\text{Var}\left[R^2 \right] \approx \text{Var} \left[ R \right]^2 $, or large $N$ such that the first term is sufficiently dominant, we can make the simplifying approximation: 
\begin{align}
\text{Var}  \left[  f_{(N)}(i,j) \right] &\approx \frac{1}{N}\sum_{{i'}=1 }^{n} \sum_{{j'}=1}^{m} f^2(i',j') = \frac{\left< f^2 \right>}{N}. \label{eq: noise approx good}
\end{align}
The variance in a finite reconstruction is independent of the distribution of $R$, depending only on the spatial sum of the square of the desired function $f$ and the number of basis members $N$. 

The local signal-to-noise ratio (SNR) is:
\begin{align} \label{eq:global SNR definition}
\text{SNR}(i,j) \equiv \frac{\text{E}[f_{(N)}(i,j)]}{\sqrt{\text{Var}[f_{(N)}(i,j)]}} \approx \frac{f(i,j)}{\sqrt{\left<f^2 \right>/ N}},
\end{align}
with the corresponding global SNR being the root-mean-square (RMS) average of the above expression:
\begin{align} \label{eq: SNR result non-orthog}
\text{SNR} \approx \sqrt{\frac{N}{nm}}.
\end{align}
This leads to a form of noise--resolution uncertainty principle:
\begin{align} \label{eq:Noise-resolution Uncertainty}
\big( \text{SNR} \big)^2 \times \big( nm \big) \approx N,
\end{align}
where $nm$ is a measure of the resolution and $N$ is the number of basis members or, in a sense, is the number of measurements required. Hence there is a direct trade off between resolution and SNR for a given number of basis members or measurements. Note that this statement  refers to the noise inherent to the process of image reconstruction from a non-orthogonal random basis as outlined above, and in no way includes the noise contributions and physical limitations of experimental realization which will have an additional noise--resolution uncertainty principle \cite{gureyev2016spatial}.

\subsection{Random-matrix basis with Gram--Schmidt orthogonalization} \label{sec: Random bases w GS}

A finite realized set of random-matrix basis members will not be strictly orthogonal, as seen in Sec.~\ref{sec: expected orthog}. They can, however, be made strictly orthonormal via the Gram--Schmidt process \cite{AntonRorres}. Further, supposing we have a set of weighting coefficients already generated for our original random basis set, we can transform both objects to achieve a more efficient representation via:
\begin{align}
\begin{split}
w'_k 	&= w_k - \sum_{\ell=1}^{k-1} \frac{\left< R_{\ell},R_{k} \right> }{\left< R_{\ell},R_{\ell} \right>} w_{\ell}, \qquad  \tilde{w}_k = \frac{w'_k }{\left< R'_k,R'_k \right>},  \\
R'_k &= R_k - \sum_{\ell=1}^{k-1} \frac{\left< R_{\ell},R_{k} \right> }{\left< R_{\ell},R_{\ell} \right>}  R_{\ell}, \qquad  \tilde{R}_k = \frac{R'_k }{\left< R'_k,R'_k \right>}.
\end{split}
\end{align}
Here, 
the primed quantities are intermediate steps and the tildes represent the orthonormalized basis quantities. 

The implementation of the Gram--Schmidt algorithm can be computationally expensive and even unstable. To improve the stability and efficiency of the algorithm, it can be implemented via QR decomposition and variants thereof \cite{daniel1976reorthogonalization, pelliccia2017practical}. We make particular note of the use of the Householder transformation, in this context \cite{Diaconis}.  Irrespective of how orthonormalization is achieved, the corresponding function reconstruction is:
\begin{align} \label{eq: function reconstruction orthog}
\begin{split}
f(i,j) & = \sum_{k=1}^{nm} \tilde{w}_k \tilde{R}_k (i,j) \\
 & \approx \sum_{k=1}^{N} \tilde{w}_k \tilde{R}_k (i,j) \equiv f^{(N)} (i,j),
\end{split}
\end{align}
where $f^{(N)}$ is the approximate version of $f(i,j)$ projected onto the incomplete $N$ member basis set ($N<nm$). Note that we have set $N$ to be less than $nm$ above since this directly implies the basis to have insufficient members to be complete; the set would be complete with $N$ members and over-complete with greater than $N$ members.
Note also that the subscript $N$ in the non-orthogonal case is swapped for a superscript in the orthonormal case, to distinguish the two. Finally, we augment the basis via inclusion of one 
constant basis member $R_1 (i,j) = 1/\sqrt{nm}~\forall~ i,j$, to improve 
convergence. That is, $R_1 (i,j)$ will efficiently achieve the spatial average of the desired matrix $f(i,j)$, and the random basis will then only be required to capture deviations from this spatial average. 

As expressed in Eq.~(\ref{eq: function reconstruction orthog}), an exact reconstruction of $f(i,j)$ is achieved with $nm$ basis members ({\em i.e.} a complete set). Seeking a more explicit statement regarding the noise present in a reconstruction using an incomplete random basis with Gram--Schmidt utilization, consider the variance of the special case consisting of a single constant value basis member:
\begin{align}
\text{Var} \left[ f^{(N)} (i,j) \right] = \frac{1}{nm} \left< \left( \bar{f} - f(i,j) \right)^2 \right> \equiv  \text{Var}[f],
\end{align}
where $\bar{f}$ is the spatial average of $f(i,j)$. Now consider the case with a complete basis set and hence an exact reconstruction. In this latter case the variance per pixel vanishes. Finally, consider an intermediate  case using an incomplete basis set with between two and $(nm-1)$ basis members inclusive. Given that the basis matrices are inherently random, each basis member will on average contribute equally to the reconstruction. This is similar in sentiment to the idea of equally likely micro-states in statistical mechanics \cite{SearsSalinger}. Adding independent deviations in quadrature implies the variance of an incomplete basis set is given by linearly interpolating between the 1 special-constant basis case and the $(nm)$ basis-set case:
\begin{align} \label{eq: var GS recon}
\text{Var} \left[ f^{(N)} (i,j) \right] = \text{Var}[f] \left( 1 - \frac{N}{nm} \right).
\end{align}
It is emphasized that the dependence on $f$ only arises due to the inclusion of the initial privileged basis member $R_1(i,j) = 1/\sqrt{nm} ~\forall ~ i,j$, to boost convergence. The above expression was seen to be consistent with the results of numerical simulation, for several different probability distributions (Gaussian, Poissonian, and uniform).

We adopt the same global SNR definition as 
in Sec.~\ref{sec:Derivation of Fn Rec w/ RB}. Hence:
\begin{align}
\text{SNR} = \sqrt{ \frac{\left<f^2 \right>}{\text{Var}[f]} \left( \frac{1}{nm-N} \right)}.
\end{align}
Note the scaling of SNR here compared to the non-orthogonal case. Previously we had $\text{SNR} \approx \sqrt{N/(nm)} $, whereas now we have $\text{SNR} \propto \sqrt{1/(nm-N)}$. Note also that ``noise'' in this context refers to spatially random errors in the synthesis of $f(i,j)$ that is obtained by superposing random matrices in an idealized setting devoid of experimental noise. Each will reach infinite SNR: the Gram--Schmidt case in $N=nm$ orthonormal random matrices, whilst the non-orthogonal case requires $N \rightarrow \infty$ basis matrices. 
The above comparison is independent of the particular random distribution employed to develop the random basis. Further, in the case of an incomplete basis set for the Gram--Schmidt case, the representation of the reconstruction is not unique and depends on the order of the basis elements being 
orthonormalized. Hence a further reduction in noise can be achieved by averaging over different Gram--Schmidt-basis reconstruction results \cite{pelliccia2017practical}.

\section{Computational Ghost Imaging Shot Noise Analysis} \label{sec:GI}


To compare conventional direct imaging and computational ghost imaging with respect to fluence or dose, we can make a preliminary assessment by counting ``balls'' in ``bins''. 
Here, the bins correspond to a pixel detector and the balls to the photons.
For the direct image, suppose we have $n \times m$ bins with $n m \tilde{\lambda}$ balls to make the image.  
We will distribute $\tilde{\lambda}$ balls into each bin and let $f(i,j)$ be the proportion of balls that will go in at each $(i,j)$ bin ({\em i.e.} $0\le f(i,j) \le 1$ is the transmission value of the sample). We describe the influence of noise by the Poisson distribution given its relevance in describing shot noise present in pixel photon measurements \cite{mandel1995optical,blanter2000shot}. This gives the normalized bin value $U(i,j)$ in the direct imaging case as:
\begin{align}
U(i,j) =  \frac{P( f(i,j) \tilde{\lambda} )}{\tilde{\lambda}},
\end{align}
where $P$ represents the Poisson distribution with a mean value of $ f(i,j) \tilde{\lambda}$. 
The expected value and variance of the normalized bin value 
is representative of the noise free value and the expected noise respectively. That is, we expect each normalized bin value to converge to:
\begin{align}
\text{E}[U(i,j)] = \text{E} \left[ \frac{P(  f(i,j)\tilde{\lambda})}{\tilde{\lambda}} \right] = f(i,j),
\end{align}
where we have used E$[P(  f(i,j) \tilde{\lambda})] =  f(i,j) \tilde{\lambda}$. 
The expected noise is given by:
\begin{align}
\text{Var}[U(i,j)] = \text{Var} \left[  \frac{P( f(i,j)\tilde{\lambda})}{\tilde{\lambda}} \right] = \frac{f(i,j)}{\tilde{\lambda}},
\end{align}
where we have used Var$[P( f(i,j) \tilde{\lambda})] = f(i,j) \tilde{\lambda}$. 
The total expected noise in the entire direct image is the spatial sum of these values:
\begin{align}
\text{Var} \left[ \text{Im}_{\text{DI}} \right] = \sum_{i=1}^{n} \sum_{j=1}^{m} \frac{f(i,j)}{ \tilde{\lambda}} = \frac{\left< f \right>}{\tilde{\lambda}}.
\end{align}

Moving onto the ghost imaging case, we now perform this comparison for computational ghost imaging. We wish to employ a scheme that will converge to the exact image in a finite number of measurements. This precludes simply using randomly generated, non-orthogonal masks which require infinitely many basis members to reach an over complete basis set. Nor will we consider taking $nm$ non-orthogonal masks and seek to orthonormalize them (via the Householder transformation or otherwise), whereby experimental uncertainties will propagate in a relatively complex fashion. Instead, as a baseline analysis, we will consider a computer generated complete set of $nm$ orthonormal random matrices $\{ \tilde{R}_{k}(i,j)\}$ of size $n \times m $ which includes one constant member $\tilde{R}_1(i,j) = 1/\sqrt{nm} ~ \forall ~i,j$. In general, these can be mapped to non-negative ghost imaging masks $\{ I_k(i,j) \}$ via the transformation:
\begin{align}
I_k (i,j) = \frac{1}{\xi} \left[ \tilde{R}_k (i,j)  - \text{min} \right],
\end{align}
where $\{ \tilde{ R}_{k}(i,j)\} \in [\text{min},  \text{max}]$, $\xi \equiv (\text{max}-\text{min})$ and $\{I_k(i,j)\} \in [0,1]$. Further, we will consider the order: mask, then sample (which is the conventional set-up for ghost imaging but not necessarily a requirement; see Sec.~VI below for further information on this point). To conserve the dose that the sample is exposed to, we will enforce that we have $\tilde{\lambda}$ balls downstream of the mask (which is consistent with $nm\tilde{\lambda}$ total balls reaching the sample). This gives the uniform incident illumination $x$ prior to the mask as:
\begin{align}
\left< I_k \right> \frac{x}{nm} &= \tilde{\lambda}, \qquad
\Rightarrow x = \frac{nm}{\left< I_k \right>} \tilde{\lambda }.
\end{align}
This construction has the associated image or normalized bin value:
\begin{align} \label{eq: norm bin ghost}
U (i,j) = \sum_{k=1}^{nm} \left( B_k + \eta B_1  \right) \tilde{ R}_k (i,j),
\end{align} 
where $B_k$ is the $k^{\text{th}}$ bucket reading and $\eta B_1$ is (min) multiplied by the spatially integrated sample transmission function $ \left<f \right>$, which can be determined from the constant basis member via $ \eta \equiv \text{min} / (1/\sqrt{nm} - \text{min})$. Explicitly, the measured bucket reading $B_k$ is given by:
\begin{align} \label{eq: norm bucket ghost}
B_k = \xi \frac{\left< I_k \right>}{\tilde{\lambda}} P \left( \frac{\tilde{\lambda }}{\left< I_k \right>} \left< I_k,  f \right> \right).
\end{align}
We now calculate the expectation and variance of the effect of Poisson noise on the ghost imaging case. The expectation is:
\begin{align}
\begin{split}
\text{E} \left[  U(i,j) \right] 
&= \sum_{k=1}^{nm} \left( \text{E} [B_k] + \eta \text{E} [ B_1 ] \right) \tilde{ R}_k (i,j) \\
&= \sum_{k=1}^{nm} \left( \xi \left< I_k,  f \right> + \text{min} \left< f \right> \right) \tilde{ R}_k (i,j) \\
&=  \sum_{k=1}^{nm} \left< \tilde{ R}_k,  f \right> \tilde{ R}_k (i,j) = f(i,j).
\end{split}
\end{align} 
For the variance: 
\begin{align}
\begin{split}
\text{Var}\left[ U(i,j) \right] &= \left( \tilde{R}_1 (i,j) + \eta \sum_{k=1}^{nm} \tilde{R}_k (i,j) \right)^2 \text{Var}[B_1] \\
&+  \sum_{k=2}^{nm} \text{Var} [B_k]  \tilde{R}_k^2 (i,j),
\end{split}
\end{align}
where we have once again set the covariance in different bucket readings to zero. The bucket reading variance is:
\begin{align} 
\begin{split}
\text{Var} [B_k] 
&= \xi^2  \text{Var} \left[  \frac{\left< I_k \right>}{\tilde{\lambda}} P \left( \frac{\tilde{\lambda }}{\left< I_k \right>} \left< I_k,  f \right> \right)  \right] \\
&= \xi^2 \frac{\left< I_k \right>}{\tilde{\lambda}} \left< I_k,  f \right>,
\end{split}
\end{align}
which implies the variance of a normalized bin filled with balls via the ghost imaging scheme to be:
\begin{align}
\begin{split}
\text{Var}\left[ U(i,j) \right] &= \left( \tilde{R}_1 (i,j) + \eta \sum_{k=1}^{nm} \tilde{R}_k (i,j) \right)^2 \xi^2 \frac{\left< I_1 \right>}{\tilde{\lambda}} \left< I_1,  f \right> \\
&+  \sum_{k=2}^{nm} \xi^2 \frac{\left< I_k \right>}{\tilde{\lambda}} \left< I_k,  f \right>  \tilde{R}_k^2 (i,j).
\end{split}
\end{align}
This in turn implies that the variance of the ghost image, being the spatial sum of bins, is:
\begin{align} \label{eq:  37}
\begin{split}
\text{Var} \left[ \text{Im}_{\text{GI}} \right] &= \left< \left( \frac{1}{\sqrt{nm}} + \eta \sum_{k=1}^{nm} \tilde{R}_k (i,j) \right)^2 \right> \xi^2 \frac{\left< I_1 \right>}{\tilde{\lambda}} \left< I_1,  f \right> \\
&+  \sum_{k=2}^{nm} \xi^2 \frac{\left< I_k \right>}{\tilde{\lambda}} \left< I_k,  f \right>,
\end{split}
\end{align}
where we have used  $\tilde{R}_1(i,j) = 1/\sqrt{nm} ~ \forall ~i,j$ and $\left< R_k^2 \right> = 1$. 
This implies that computational ghost imaging may reduce dose over direct imaging by boosting SNR \cite{zhang2018tabletop} if a basis set exists such that for the same image, the variance of the ghost image is less than that of the direct image.  Such dose reduction therefore requires the following inequality to be satisfied:
\begin{align}
 \Omega \xi^2 \left< I_1 \right> \left< I_1,  f \right> + \xi^2 \sum_{k=2}^{nm}  \left< I_k \right> \left< I_k,  f \right> < \left< f \right>,
\end{align} 
where $\Omega \equiv \left< (1/\sqrt{nm} + \eta \sum_{k=1}^{nm} \tilde{R}_k (i,j) )^2 \right> $. Conversely, if a basis set is such that the above inequality is violated, the ghost imaging case will be less dose efficient than the corresponding direct-imaging case ({\em cf.}~a similar inequality obtained in a different context, recently reported by Gureyev {\em et al.}~\cite{GureyevGhost2018}).
 
It is reasonable in some cases to make the approximation $\sum_{k=1}^{nm} \tilde{R}_k \approx 0 ~ \forall ~ i,j$ which reduces Eq.~\ref{eq:  37} to:
\begin{align}
\text{Var} \left[ \text{Im}_{\text{GI}} \right] & \approx  \sum_{k=1}^{nm} \xi^2 \frac{\left< I_k \right>}{\tilde{\lambda}} \left< I_k,  f \right>.
\end{align}
Ignoring $\xi^2$ and $\left< I_k \right>$, which scale approximately as $1/nm$ and $nm$ respectively, suggests that in order for ghost imaging to reduce dose, the sum of bucket readings needs to be less than the spatial sum of the transmission function. That is, we want the mask-sample combination to be more absorbing than the sample alone which corresponds to fewer balls being detected, which when normalized, produces a lower uncertainty.

As an illustration of the logical possibility for dose reduction on account of the above inequalities, consider the following extreme limiting case.  Suppose one wishes to perform computational ghost imaging on a sample whose two-dimensional intensity transmission function is a pixellated grayscale image of a given highly textured structure.  Suppose too that a very small number of pseudo-random illuminating masks is used, each of which happen to have an unusually high degree of correlation with the illuminated sample.  The resulting computational ghost image will then be obtainable with an unusually small number of bucket measurements, and a correspondingly reduced dose.  While this scenario is of course very highly improbable when genuinely random masks are used, the possibility of such a scenario becomes significantly more likely when one utilizes feedback computational ghost imaging, a point which is very briefly explored later in the present paper.      

It is physically reasonable that the dose-reduction inequalities should be both object- and mask-dependent.  Notwithstanding this fact, if one wishes to image a particular class of object, it may be useful to average the above dose-reduction inequalities over a statistical ensemble of possible objects.

The preceding analysis regarding dose is conservative as it does not consider incorporation of compressive sensing ideas \cite{Qaisar2013}. That is, compressive sensing allows one to perform ghost imaging in less than $nm$ measurements, which is beneficial regarding reduced dose. Moreover, in cases where the above inequality is violated, the necessity for a more sophisticated approach to ghost reconstruction---such as given by compressive ghost imaging methods \cite{katz2009compressive, zerom2011entangled, katkovnik2012compressive,
yu2014adaptive}---is made apparent as a necessary condition for ghost imaging to reduce dose when compared to its direct-imaging counterpart \cite{GureyevGhost2018}.

\section{X-ray Phase Contrast Computational Ghost Imaging: Theory} \label{sec: PCGI}


Here we apply the random-matrix-basis concept to x-ray phase contrast computational ghost imaging.  In this context, phase contrast may be defined as any imaging modality yielding intensity maps that are a function of the phase of the field that is input into the associated imaging system \cite{paganin2006coherent}.  In an x-ray setting, phase contrast permits visualization of x-ray-transparent structures.  While many means for realizing x-ray phase contrast exist in a direct-imaging setting \cite{paganin2006coherent}, there is a dearth of such methods for x-ray ghost imaging.  Indeed, there are only a handful of papers on x-ray ghost imaging \cite{Yu2016,pelliccia2016experimental,schori2017,zhang2018tabletop,pelliccia2017practical}, all of which demonstrate absorption-contrast x-ray ghost imaging, which is only sensitive to the magnitude of the field visualized by the associated ghost imaging system.  While means exist for realizing phase contrast in non-x-ray ghost-imaging settings \cite{shirai2011ghost,ghostphase1,ghostphase2,ghostphase3}, these are not readily transferred to x-ray imaging.        

\begin{figure*}
\includegraphics[width=0.8\textwidth,scale=0.12]{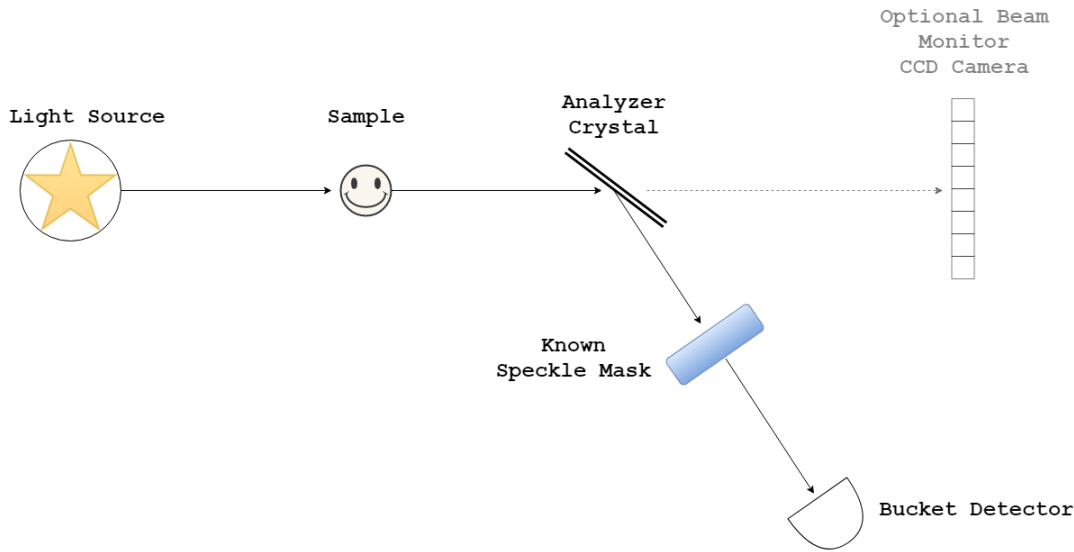}
\caption{Schematic of computational x-ray phase contrast ghost imaging. }
\label{fig: PCGI Set Up)}
\end{figure*}

The proposed experimental set-up for computational x-ray phase contrast ghost imaging 
utilizes a quasi-monochromatic \cite{mandel1995optical} x-ray source, an analyzer crystal \cite{Forster1980}, an ensemble of known speckle masks and a bucket detector---see Fig.~\ref{fig: PCGI Set Up)}. The x-ray phase information imparted by the refractive properties of the sample will be encoded as transverse spatial intensity variations via the Fourier-space filtration provided by reflection from the analyzer crystal \cite{paganin2006coherent}. Indeed, the angle-dependent reflectivity function (rocking curve) of the analyzer crystal is key in this setup to achieving bucket signals which are sensitive to the phase gradients of the impinging x-ray radiation.  The analyzer-crystal-filtered intensity field is then incident on a known speckle mask and the combined effect on the intensity is integrated up in the bucket detector. The process is repeated for each speckle mask in the set of known masks. Note that in an experiment one can obtain such a set of speckle masks by transversely scanning a single larger mask \cite{schori2017}.  Note also that for each bucket-signal acquisition the beam fully illuminates the sample. One may optionally include position-sensitive beam monitors at either or both of the indicated positions, to adapt to inherent temporal fluctuations of the source from measurement to measurement. 

Consider the intensity $I(i,j)$ and phase $\phi(i,j)$ of a beam measured in a plane perpendicular to the beam propagation ($z$-axis) on a pixel grid $(i,j)$. We consider the case that the sample and mask are each made from a single and in-general different uniform material.  This single-material assumption is not particularly restrictive regarding the mask, since the sole purpose of the mask is to generate a set of linearly independent spatially random intensity patterns.  Further, this assumption is far less restrictive than it might appear, regarding the class of objects that may be imaged, for two reasons: (i) the same assumption has been successfully employed in several hundred papers in a propagation-based phase contrast x-ray tomography setting, since many samples of interest may be locally ({\em i.e.} in three spatial dimensions) described as consisting of a single material of interest \cite{paganin2002,SNRboost1,SNRboost2}; (ii) for sufficiently high x-ray energies, low-atomic-number materials such as those in soft biological tissues, behave approximately as a single material, namely electrons \footnote{P.~Cloetens, private communication, 2009.}, with the atomic nuclei having only a small effect on both the absorptive and refractive properties of the stated soft-tissue materials in this regime \cite{WuLiuYan2005}.  
 
Under both the single-material assumption and the projection approximation, the phase shift due to the sample is \cite{paganin2006coherent}:
\begin{align}
\phi_{\text{sample}} (i,j) &= - k \delta T(i,j),
\end{align}
where $k = 2 \pi/ \lambda$ is the wavenumber, $\lambda$ is the wavelength, $\delta \equiv 1 - \mathfrak{R} \{n\} $, $\mathfrak{R} \{n\}$ is the real part of the refractive index of the sample material and $T(i,j)$ is the projected thickness of the sample. The angular deviation from the $z$-axis, which is taken to be the optic axis, post the sample is:
\begin{align}
\Delta \theta(i,j) =-\frac{1}{k}  \frac{\partial}{\partial x} \bigg[ \phi_{\text{sample}} (i,j)  \bigg] 
=  \delta  \frac{\partial T}{\partial x} .
\end{align}
From here we can calculate the intensity of the beam post the analyzer crystal interaction. We use the Beer--Lambert law of absorption \cite{paganin2006coherent} to model the sample interaction (and mask interaction). The reflectivity of the crystal as a function of deflection angle, namely the rocking curve \cite{Forster1980}, is taken to be known. Assuming a linear approximation to the rocking curve \cite{Chapman1997} gives: 
\begin{align}
\nonumber I_{\text{sample}}(i,j)  &= I_0 \exp[-\mu T(i,j)][ \alpha + \beta \Delta \theta(i,j) ] \\
&= I_0 \exp[-\mu T(i,j)] \left( \alpha + \beta \delta  \frac{\partial T}{\partial x} \right),
\end{align}
where $I_0$ is the uniform intensity input, $\mu$ is the linear attenuation coefficient of the sample, $\alpha$ is the zero angle deflection reflectivity and $\beta$ is a linear approximation to the rocking curve slope. We then subject the intensity field to absorption from the speckle mask, to give the bucket detector reading $B_k$ as:
\begin{align}
B_k = \left< I_0 \exp \bigg[ -\mu T(i,j) -\mu_0 R_k(i,j) \bigg] \left( \alpha + \beta \delta  \frac{\partial T}{\partial x} \right) \right>,
\end{align}
where $\mu_0$ and $ R_k(i,j)$ are the absorption coefficient and projected thickness of the mask respectively. Note, in this context and for the remainder of the paper, $R_k(i,j)$ is a random value greater than zero which is slightly different than in the previous two sections where it was a zero centered random value.

We can compute the x-ray phase contrast computational ghost image (PCCGI) using the standard ghost imaging formula (Eq.~(\ref{eq: Ghost Image Formula})), or via the Gram--Schmidt process. For the standard ghost imaging case, we will require a normalization of $ \text{Var} \left[ I_0  \exp (- \mu_0 R)  \right]$ to obtain the expected image:
\begin{align}
\text{E} \left[ \text{PCCGI}(i,j)  \right]  = \alpha    \left( 1 - \frac{ \beta \delta}{\mu \alpha}  \frac{\partial}{\partial x}  \right)  \exp [ - \mu T].
\end{align}
The expected noise in the image will be given by the variance, 
\begin{align}
\begin{split}
\text{Var} &\left[ \text{PCCGI}_{(N)} (i,j) \right] \\
&\approx \frac{\left<  \left[ \alpha    \left( 1 - \frac{ \beta \delta}{\mu \alpha}  \frac{\partial}{\partial x}  \right)  \exp ( - \mu T )\right]^2 \right> }{N},
\end{split}
\end{align}
or with Gram--Schmidt utilization:
\begin{align}
\begin{split}
\text{Var} \left[ \text{PCCGI}^{(N)} (i,j) \right] \quad\quad\quad\quad\quad\quad\quad\quad\quad\quad\quad\quad\\
\approx \text{Var} \left[ \alpha    \left( 1 - \frac{ \beta \delta}{\mu \alpha}  \frac{\partial}{\partial x}  \right)  \exp ( - \mu T ) \right] \left( 1 - \frac{N}{nm} \right),
\end{split}
\end{align}
where the variance of $ \alpha    \left( 1 - \frac{ \beta \delta}{\mu \alpha}  \frac{\partial}{\partial x}  \right)  \exp ( - \mu T) $ is relative to the spatial average.


The expected value of the phase contrast ghost image has the form:
\begin{align}
\text{Image}_{\text{PCCGI}}= C \left( 1 - G \frac{\partial}{\partial	x} \right) \exp \left[- \mu T(x,y) \right],
\label{eq: ExpectedPXGI}
\end{align}
where $C=\alpha$ and $G=\beta\delta/\mu\alpha$ are known real constants. This x-ray phase contrast ghost image can be viewed as encoding quantitative information regarding the projected thickness $T$ of the sample, thereby inviting solution to the associated inverse problem \cite{Sabatier2000} of obtaining $T$ from $\text{Image}_{\text{PCCGI}}$.  The situation is somewhat analogous to Gabor's original conception of (inline holographic) imaging as a two-step process, namely recording followed by reconstruction \cite{Gabor}.  To this end, Fourier transform the above differential equation with respect to $x$, use the Fourier derivative theorem, then solve the resulting algebraic equation for the Fourier transform of $\exp(-\mu T)$.  Inverse Fourier transformation gives an expression mathematically identical to that previously obtained by Paganin {\em et al.} in a direct-imaging x-ray phase-contrast context unrelated to ghost imaging \cite{paganin2004,briedis2005,vine2007}:
\begin{align}
T(x,y) = - \frac{1}{\mu} \ln \left\{ \mathfrak{F}_x^{-1} \left[ \frac{\mathfrak{F}_x \left( \text{Image}_{\text{PCCGI}} \right)}{C \left( 1 - i G k_x \right)} \right] \right\}.
\label{eq:XPCGI_Reconstruction_formula}
\end{align}
Here, $\mathfrak{F}_x$ and $\mathfrak{F}_x^{-1}$ denote the Fourier transform and inverse Fourier transform in the $x$ direction respectively, and $k_x$ denotes the Fourier-space coordinate dual to $x$.  We use a Fourier-transform convention in which the Fourier derivative theorem takes the form $\partial/\partial x = \mathfrak{F}_x^{-1}(ik_x)\mathfrak{F}_x$ \cite{paganin2006coherent}. For $G \ne 0$, the Fourier-space filter $1/(1-iGk_x)$ has modulus $(1+G^2k_x^2)^{-1/2}$ that never vanishes, never has division-by-zero instability, and never exceeds unity.  This class of inverse-problem solution belongs to the category of SNR-boosting algorithms recently investigated by Gureyev {\em et al.}~\cite{gureyev2017unreasonable}.  A related member of this algorithm class \cite{paganin2002} has been shown to boost SNR by factors $F$ on the order of hundreds (or, equivalently, enable dose and data acquisition time to be reduced by factors $F^2$ on the order of tens of thousands) in the process of reconstruction \cite{SNRboost1,SNRboost2,SNRboost3,SNRboost4,SNRboost5}. 


\section{X-ray Phase Contrast Computational Ghost Imaging: Simulations}

\subsection{Parameters and rocking-curve model}

In our simulations, the analyzer-crystal rocking curve is approximated by the Pearson VII distribution \cite{PearsonVII}:
\begin{align}
R(\theta) = R_0 \left( 1 + \frac{\theta^2}{\mathfrak{M} a^2} \right)^{-\mathfrak{M}}.
\end{align}
Here, $R$ is reflectivity, $R/R_0$ is relative reflectivity and the independent variable $\theta$ is the deviation from the Bragg angle associated with the operative analyzer-crystal reflection; $R_0$, $\mathfrak{M}$ and $a$ are parameters of the crystal and the x-ray energy. Note that the same symbol is used for both reflectivity and a random variable; it will always be clear from the context which meaning is intended. Taking a Taylor series about $\theta_0$, corresponding to a single chosen angular detuning of the analyzer crystal from the Bragg condition for which the entirety of the images is taken, we obtain an expression for the linearization parameters that approximate the rocking curve ($\alpha$ and $\beta$) in terms of $R_0$, $\theta_0$, $\mathfrak{M}$ and $a$:
\begin{align}
\alpha 	&= R(\theta_0) =  R_0 \left( 1 + \frac{\theta^2_0}{\mathfrak{M} a^2} \right)^{-\mathfrak{M}}, \\
\beta 	&=  \frac{dR}{d\theta}\bigg|_{\theta_0} = - \frac{2\theta_0 R_0}{a^2} \left( 1 + \frac{\theta^2_0}{\mathfrak{M} a^2} \right)^{-\mathfrak{M}-1}.
\end{align}
The numerical parameters for the Pearson VII curve correspond to a Si(333) analyzer crystal at illuminating energy $E=40$keV, and were taken from Majidi {\em et al.}~\cite{majidi2014noise} as $a =0.7146$, $\mathfrak{M} =2.3737$, $R_0=1$. The uniform initial intensity was assumed to be normalized, so that $I_0 = 1$. 
Material values were obtained from the NIST database \footnote{Attenuation coefficients taken from: https://physics.nist. gov/PhysRefData/XrayMassCoef/tab3.html. Form factor $f_1$ taken from: https://physics.nist.gov/ PhysRefData/FFast/html/form.html. Molar mass and density values taken from: https://physics.nist.gov/PhysRefData/XrayMassCoef/ tab1.html.} and are tabulated below. 

\begin{center}
\begin{tabular}{ |c|c|c|c|c|c| }
\hline
 \multicolumn{6}{|c|}{Material Parameters for $E \approx 40$keV } \\
 \hline
 Element 	& Z 	& $M_A$[g/mol] 	& $\mu / \rho$[cm$^2$/g]& $\rho$[g/cm$^3$] 	& $f_1$ \\
 \hline
Carbon 		& 6 	& 12.011 		& $0.2076$				& 1.700  			& 6.00115\\ 
 Aluminium 	& 13 	& 26.982 		& $0.5685$				& 2.699  			& $13.0206$\\ 
 Copper 	& 29 	& 63.546 		&  4.862				& 8.960  			& $29.2497$\\
Gold 	 	& 79 	& 196.966		& $12.98$				& $19.32$ 			& $79.1108$ \\ 
 \hline
\end{tabular}
\end{center}
Note, form factor $f_1$ values were obtained at $E =39.19543$ keV. We hence computed the refractive index decrement $\delta$ via \cite{attwood2017x}:
\begin{align}
\delta = \frac{n_a r_0 \lambda ^2 }{2 \pi} f_1,
\end{align}
where $r_0$ is the classical electron radius, 
$\lambda$ is the x-ray wavelength, $n_a$ is:
\begin{align}
n_a \approx \frac{\rho N_A}{M_A},
\end{align}
$\rho$ is the material density, $N_A$ is Avogadro's number 
and $M_A$ is the molar mass. 

\subsection{Idealized x-ray phase contrast computational ghost imaging simulations}

In the present subsection we make two natural simplifying assumptions, both of which will be dropped in the next subsection. (i)
The idealized ghost imaging case we consider here has a 1:1 pixel-to-speckle correspondence, giving a unique random illuminating intensity at each pixel.  This amounts to seeking a ghost reconstruction with resolution equal to the mask speckle size, a natural restriction that is consistent with the fact that the point-spread-function of the reconstruction is equal to the auto-covariance of the ensemble of speckle masks \cite{ferri2010, pelliccia2017practical,GureyevGhost2018}. Indeed, the 1:1 pixel-to-speckle correspondence investigated here is optimal, since sampling the speckles any more finely does not improve spatial resolution (in the ghost reconstruction) beyond the maximum resolution given by the speckle size. (ii) The effects of detector noise are not considered in the present subsection.  Any  deleterious effect of detector noise in the present context is typically rather mild, since bucket signals are rather insensitive to noise on account of the fact that they integrate over the entire transverse extent of the beam. 

The simulated sample being imaged is an amorphous carbon ellipsoid with projected thickness:
\begin{align}
T(x,y) = 2 \mathfrak{R}\big[\sqrt{ (r/\aleph)^2-(y/\aleph)^2-(x/\aleph)^2}\big],\label{eq:EllipsoidProjectedThickness}
\end{align}
where $\aleph = 8$, $r=16$mm and $\mathfrak{R}$ denotes the real part.  For this simulated object, $R/R_0$ was approximately within the range (0.3, 0.65), consistent with the linear approximation to the rocking curve that has been adopted. Each pixel is 1mm$\times$1mm in size, corresponding to an overall field of view on the order of several centimeters, as is often used {\em e.g.} in medical-imaging x-ray synchrotron beamlines \cite{stevenson2010first} and laboratory-source x-ray imaging \cite{vine2007}.

We will compare image reconstructions of the projected thickness performed with the standard ghost imaging formula (Eq.~(\ref{eq: Ghost Image Formula})) and with Gram--Schmidt utilization (Eq.~(\ref{eq: function reconstruction orthog})). The Pearson VII rocking curve linearization is performed about the 50\% relative reflectivity point to obtain a maximum range of the approximately linear region, which was taken to span from 0.2--0.8. The linear-approximation parameters were calculated to be $\alpha = 0.5011$ and $\beta = 9.3875 \times 10^{5}$/($\mu$rad). 

\begin{figure*}
\includegraphics[width=\textwidth,scale=0.35]{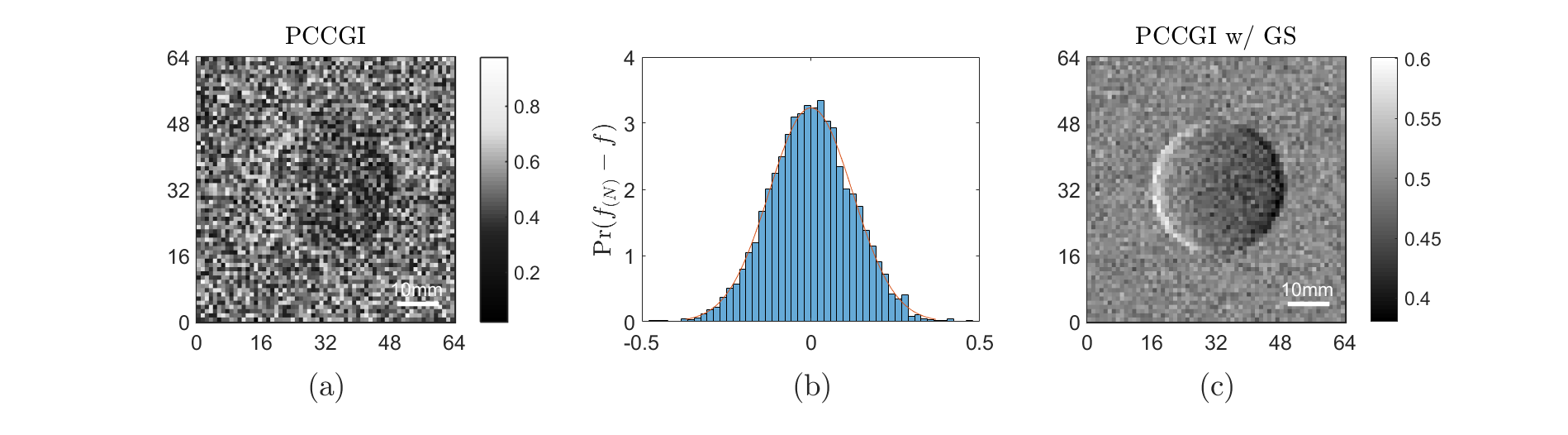}
\caption{Idealized x-ray phase contrast computational ghost imaging simulation of a carbon ellipsoid on a 64 by 64 millimeter pixel array. (a) Reconstruction performed with standard ghost imaging formula with $N=16nm$. (b) Pixel variation from expected value in (a) as a probability distribution, overlaid with predicted distribution using predicted variance and assumed Gaussian statistics. (c) Reconstruction performed with Gram--Schmidt utilization for $N=0.5nm$.}
\label{fig:ideal x-ray PCCGI comparison w GS}
\end{figure*}

The results of this first simulation are given in Fig.~\ref{fig:ideal x-ray PCCGI comparison w GS}. Figure~\ref{fig:ideal x-ray PCCGI comparison w GS}(a) exhibits x-ray ghost differential phase contrast, as the white edge bounding the left side of the ellipsoid and the corresponding dark edge on the right side of the ellipsoid.  This corresponds to the derivative term in Eq.~(\ref{eq: ExpectedPXGI}), which in turn arises from the refractive properties of the sample. Note that, while these images ({\em e.g.} Fig.~\ref{fig:ideal x-ray PCCGI comparison w GS}(a)) closely resemble those obtained using analyser-crystal x-ray phase contrast imaging \cite{Forster1980}, we emphasize that Fig.~\ref{fig:ideal x-ray PCCGI comparison w GS}(a) is an x-ray {\em ghost} phase contrast image whose phase contrast is never directly measured with a position-sensitive detector. Note also that, while the eye typically perceives some absorption within the boundaries of the ellipsoid in Fig.~\ref{fig:ideal x-ray PCCGI comparison w GS}(a), this is an optical illusion related to the physiology of the human eye \cite{Mach1,Mach2}.  Thus the dominant contrast mechanism in Fig.~\ref{fig:ideal x-ray PCCGI comparison w GS}(a) is differential phase contrast, with absorption contrast being negligible. The histogram of noise present in Fig.~\ref{fig:ideal x-ray PCCGI comparison w GS}(a) is shown in Fig.~\ref{fig:ideal x-ray PCCGI comparison w GS}(b) with the numerically obtained distribution (blue bars) overlaid with a red curve based on the theory developed above.  Finally, Fig.~\ref{fig:ideal x-ray PCCGI comparison w GS}(c) shows the differential x-ray phase contrast ghost image obtained when the Gram--Schmidt process is used to orthogonalize the speckle fields prior to the ghost reconstruction. 
The significant improvement in efficiency of the reconstruction with Gram--Schmidt utilization is consistent with previous studies on absorptive x-ray phase contrast ghost imaging \cite{pelliccia2017practical}.

\subsection{Comparison of computational x-ray absorption and phase contrast ghost imaging simulations}

\begin{figure*}
\includegraphics[width=\textwidth,scale=0.4]{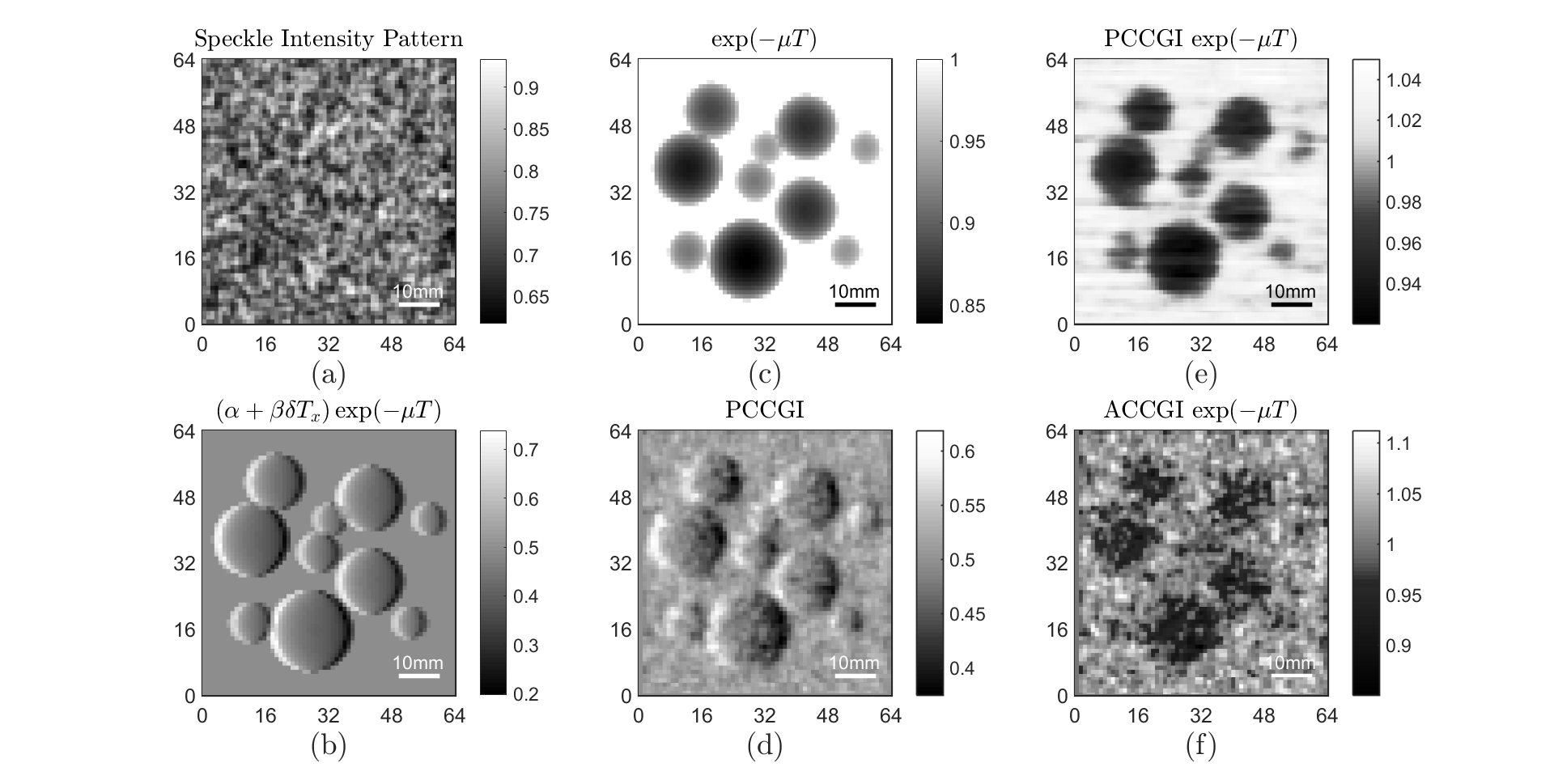}
\caption{Comparison of x-ray phase contrast and absorption contrast computational ghost imaging simulation made on a 64 by 64 millimeter pixel array with smoothed speckle intensity patterns and ``shot noise'' inclusion. Performed with $N=64nm$, $\tilde{\lambda} = 10^8$. (a) Typical speckle intensity pattern. (b) Expected phase contrast image, with $T_x \equiv \partial T/\partial x$. (c) Expected absorption contrast image.  (d) Obtained phase contrast image.  (e) Reconstructed absorption image from phase contrast. (f) Obtained absorption contrast image.  Note the boost in SNR in passing from (f) to (e).}
\label{fig:x-ray GI comparison}
\end{figure*}

To perform a more realistic x-ray ghost imaging simulation, for the case of both absorption contrast and phase contrast, we drop both of the simplifying assumptions listed at the beginning of the previous subsection.  We thus expand considerations to include spatially smoothed speckles with Poisson shot noise \cite{mandel1995optical,blanter2000shot} in the bucket reading acquisition. For the Poisson noise associated with the bucket detector as a single-pixel detector, we use the model:
\begin{align}
B_k'' = \frac{P \left( B_k  \tilde{ \lambda}' \right) }{\tilde{\lambda}'},
\end{align}
where $B_k''$ and $B_k$ are the measured and ideal bucket readings respectively, $\tilde{\lambda'}= \tilde{\lambda} \left< I_k \right>$, $\tilde{\lambda}$ is the number of ``balls'' or imaging quanta in the uniform beam, $P$ is the Poisson distribution 
and $I_k(i,j) = I_0 \exp[-\mu_0 R_k(i,j)]$ is a speckle intensity pattern. The speckle mask in this case is generated by taking a single large array of uniformly and independently distributed random numbers (0, 250$\mu$m) and filtering it with a normalized two-dimensional rotationally-symmetric  Gaussian kernel having a standard deviation of 2/3 of a pixel. Each speckle intensity pattern is then taken to  correspond to a random transverse translation of this single larger pattern. A typical speckle intensity pattern created via this method is given in Fig.~\ref{fig:x-ray GI comparison}(a). The sample being imaged here is a collection of 10 carbon ellipsoids, each described by Eq.~(\ref{eq:EllipsoidProjectedThickness}), 
%
with $\aleph = 4$ and $r=\{4,4,4,5,5,7,8,8,9,10\}$mm. 

For the case that we have more pixels than speckles, the resolution and reconstruction noise of the ghost image will not be determined by the number of pixels, but rather the number of speckles. Note that this limitation, which is always the case for the means of ghost imaging that has been considered here, may be removed or reduced if {\em e.g.} additional {\em a priori} knowledge such as sparsity constraints are able to be employed through compressive sensing methods.  Note also that the noise mentioned here is that inherent to the reconstruction process utilizing spatially random masks, which is distinct from noise obtained from experimental measurement. 

The results of this more realistic simulation are in Fig.~\ref{fig:x-ray GI comparison}.  (a) A typical illuminating speckle field is given, with (b) the expected phase-contrast ghost image (see Eq.~(\ref{eq: ExpectedPXGI})) and (c) the expected absorption-contrast image.  (d) The simulated x-ray differential phase contrast ghost image shows that, for weakly absorbing samples, the phase contrast image can still consist of quite prominent features dominated by the edge contrast associated with the spatial derivative in Eq.~(\ref{eq: ExpectedPXGI}). This x-ray ghost phase contrast leads to better visibility through the noise in (d) and ultimately retrieves (using Eq.~(\ref{eq:XPCGI_Reconstruction_formula})) a better absorption image ((e) compared to (f)). 

Whilst the simulation results required a large number $N$ of speckle intensity patterns, recall that this is a base case employing the standard ghost imaging formula (Eq.~(\ref{eq: Ghost Image Formula})).  Such a reconstruction only improves in SNR proportional to $\sqrt{N}$.  More sophisticated reconstruction algorithms---such as those that utilize compressive sensing,  machine learning, basis orthogonalization {\em etc}.---may achieve much better efficiency.  Such improvements are beyond the scope of the simulated x-ray study given in the present paper, whose core focus is on establishing proof-of-concept for x-ray phase contrast ghost imaging.  

\section{Discussion}

By considering a random matrix reconstruction of a desired two-dimensional discretized function, we have obtained $\text{SNR} \approx \sqrt{N/nm}$, which is consistent with the findings of ghost imaging studies \cite{erkmen2009signal,bromberg2009ghost,clemente2012single}. This suggests ghost imaging to be the experimental realization of the mathematical concept of a random basis reconstruction and is not inherently reliant on any underlying physical phenomena.

A key theme of this paper is employing a set of random matrices as a mathematical basis that can be used to expand discretized functions of interest. While the focus has been on expanding discretized two-dimensional functions using two-dimensional random matrices, the idea can be implemented in any integer number of dimensions. Indeed, the concept of a random-matrix basis can be considered in a purely mathematical setting, if desired, to sit alongside the Fourier series, Taylor series and other classic series as a useful tool that allows one to approximate signals as a superposition of simpler functions.  

A peculiarity of the random basis, when compared to deterministic bases, is that the spatially random basis synthesizes signals by superposing noise.  This fact facilitates applications where the requisite random matrices are spontaneously realized through (i) classical processes ({\em e.g.} via transmission of plane waves through spatially random screens in conventional x-ray ghost imaging \cite{Yu2016,schori2017,zhang2018tabletop,pelliccia2017practical} or atmospheric shifts in the case of ground based telescopes which would build on the atmospheric turbulence ghost imaging work of \cite{shi2012adaptive}); (ii) quantum processes ({\em e.g.} via photon shot noise obtained via synchrotron radiation from individual electron bunches \cite{pelliccia2016experimental}). Another peculiarity of random bases is that every basis member is in some sense statistically equivalent, implying that the basis elements cannot be ordered in a manner analogous to the ordering of deterministic bases in terms of increasing spatial frequency, rapidity of oscillation {\em etc}.  The latter point invites the previously suggested parallel between random-matrix basis members, and the thermodynamic concept of a micro-state.  Indeed, averaging over micro-states in thermodynamic ensembles is conceptually rather close to the concept of ensemble averaging over a random-matrix basis. Further exploration, of the connection between thermodynamic concepts and the concept of random-matrix bases, would form an interesting avenue for further investigation. 

We do not claim that the particular random-matrix synthesis and decomposition methods in the present paper are the most efficient.  For example, as previously stated, the efficiency and robustness of our proof-of-concept x-ray phase contrast computational ghost imaging procedure, could certainly be improved via the use of concepts and techniques from compressed sensing \cite{Qaisar2013}.  Another avenue for improvement, inspired by some recent works on inverse imaging problems, is the utilization of machine learning and artificial neural networks (see {\em e.g.} \cite{kemp2017propagation}) to realize more robust and efficient ghost-imaging algorithms that are able to make fuller use of all available {\em a priori} knowledge.  Indeed, the improvement afforded by such extensions could be quantified by comparison with the analytical ``base case'' results developed in the present paper. 

With respect to phase contrast or holographic x-ray ghost imaging, there exists the additional hurdle of potentially unwanted phase information being imparted by the mask. Recall the set-up for x-ray phase contrast computational ghost imaging (Sec.~\ref{sec: PCGI}); the re-ordering of the mask and sample, with respect to the usual sequence in which the mask occurs upstream of the sample, was enacted for this reason. Supposing we had the reverse order for mask and sample, the bucket reading would be:
\begin{align}
\begin{split}
B_k = \sum_{i=1}^{n} \sum_{j=1}^{m} I_0 \exp[-\mu T(i,j) -\mu_0 R_k(i,j)] \times \\
\left[ \alpha + \beta \delta  \frac{\partial T(i,j)}{\partial x} + \beta \delta_0 \frac{\partial R_{k}(i,j)}{\partial x}\right],
\end{split}
\end{align}
where symbols with subscript 0 are mask parameters. The spatial gradient of the mask in this case presents an unwanted contribution that corrupts the bucket reading and ghost imaging process in a manner problematic to remedy. Hence we recommend that for x-ray phase contrast and holographic ghost imaging investigations in general, moving the mask/SLM (spatial light modulator) to downstream of the sample may produce a more practical implementation. 

In a similar fashion, consider now environmental or turbulence robustness \cite{shi2012adaptive,turbulence1,turbulence2,turbulence3,turbulence4}.  The spatial distribution of the mask intensity pattern must be approximately unchanged when illuminating the sample (or {\em vice versa}) to avoid corruption of the ghost imaging process, but need not necessarily be preserved for the entire path from sample to detector (as is consistent with the findings of \cite{le2017underwater}). That is, from the perspective of a series expansion, the weighting coefficient or bucket reading must be of the form of a known basis, multiplied by sample, and spatially integrated. This permits arbitrary scrambling of the information post the mask--sample interaction.



In the case of direct imaging, the minimum resolvable area is limited by the pixel size. In the case of ghost imaging, the limiting factor is the mask speckles, which presents a potentially easier manufacturing obstacle to overcome. That is, a potential avenue of future work could be to nano-manufacture the mask, allowing ghost imaging to be performed on the sub micrometer scale.  This may yield a means to produce geometrical super resolution \cite{milanfar2010super} images ({\em cf.} \cite{ferri2005high}). 

Lastly, we consider the concept of feedback computational ghost imaging. This utilizes information gained from each speckle-field interaction with the sample to inform future intensity field inputs \cite{Assmann, Sun2016}. Thus the $n^{\text{th}}$ illuminating speckle field is adaptively chosen based on all preceding $n-1$ speckle fields and their associated bucket signals, in a manner that seeks maximal new information in each subsequent measurement.  The aim of this method is to produce a rapidly converging imaging system that seeks to reduce dose by reducing the number of measurements required---this is on-the-fly compressive acquisition, as suggested by A{\ss}mann and Bayer \cite{Assmann}. For example, we could use the gained information to essentially inform a guess of what the exact image is. Alternatively, since SNR can have a spatial distribution, we might choose to focus future intensity patterns based on real time SNR distribution results. 
Regarding implementation of this concept, and further to the recent results of Sun {\em et al.} \cite{Sun2016}, suppose one wanted to converge to the exact image via 
a global optimization algorithm. A goal could be to meet the condition that the SLM and the sample provide the same attenuation:
\begin{align}
\frac{ B_k'}{\left< I_k^2 \right> } = 1,
\end{align}
where $B_k'=\left<I_k,f \right>$ subject to appropriate normalization. This condition could be augmented into something suitable for improvement via simulated annealing \cite{kirkpatrick1983optimization}. Further, we might hope to utilize deep learning algorithms and artificial intelligence (AI) to improve convergence. Work in the area of AI imaging has dramatically progressed in recent times \cite{kemp2017propagation}, including in the ghost imaging area \cite{shimobaba2018computational}.

\section{Conclusion}

Synthesis of functions sampled on a discrete Cartesian lattice, via a linear combination of random-matrix basis elements, was considered.  This may be viewed as superposing discrete noise maps in order to approximate arbitrary discrete signals.  While this concept has broad applicability in physics, focus was given to functions of two spatial variables.  In this context, statements were made regarding orthogonality, completeness, arbitrary function reconstruction and SNR. 

Particular focus was given to applications in ghost imaging and computational imaging.  
This aligns with the idea that ghost imaging is the experimental realization of the mathematical process of expressing an image as a weighted sum of known intensity patterns (random or otherwise). 
We constructed an inequality that must be satisfied if the conventional non-compressive form of ghost imaging, which may be directly derived from the random-matrix-basis concept, is to yield reduced dose as compared to direct imaging. This inequality stems from comparing the expected noise or variance of each imaging process for a given dose. That is, {\em e.g.} if a ghost image yields a lower expected noise for a given dose than the direct image, that implies the ghost image could have achieved the same noise as the direct image with a lower dose. Cases in which this inequality is not satisfied may be viewed as motivating the need for compressive sensing or related concepts, as a necessary condition for dose reduction in a ghost imaging or computational imaging context. 

We proposed an experimental set-up to achieve computational x-ray phase contrast ghost imaging.  This was motivated by the dearth of x-ray ghost imaging experiments able to yield phase contrast, the infancy of the field of x-ray ghost imaging, and the renaissance in  coherent direct x-ray imaging that is currently underway due to phase contrast.  Both the forward and inverse problems were considered.  The former demonstrated the feasibility of x-ray differential phase contrast ghost imaging.  The latter gave a solution to the inverse problem of obtaining quantitative projected-thickness information regarding a sample, using a single x-ray differential phase contrast ghost image of that sample. 



\section*{Acknowledgements} 

We acknowledge useful discussions with Mario Beltran, Carsten~Detlefs, Jean-Pierre~Guigay, Timur~Gureyev, Andrew~Kingston, Alex~Kozlov, Kieran~Larkin, Anthony~Mays, Kavan~Modi, Glenn~Myers, Margie~Olbinado, Timothy~Petersen, Daniele~Pelliccia, Alexander~Rack, Imants~Svalbe and Tapio~Simula.  DMP acknowledges financial support from the European Synchrotron Radiation Facility.

\bibliographystyle{unsrt}
\bibliography{Bibtex_Thesis}

\end{document}